# Lowering the Barrier to Reuse through Test-Driven Search

Werner Janjic, Dietmar Stoll, Philipp Bostan, Colin Atkinson
*Chair for Software Technology, University of Mannheim*
{janjic, stoll, bostan, atkinson}@informatik.uni-mannheim.de

## Abstract

*Dedicated software search engines that index open source software repositories or in-house software assets significantly enhance the chance of finding software components suitable for reuse. However, they still leave the work of evaluating and testing components to the developer. To significantly change the risk/cost/benefit tradeoff involved in software reuse, search engines need to be supported by user friendly environments that deliver code search functionality, non-intrusively, right to developers' fingertips during key software development activities and significantly raise the quality of search results. In this position paper we describe our attempt to realize this vision through an Eclipse plug-in, Code Conjurer, in tandem with the code search engine, merobase.*

## 1. Introduction

The vision of systematically assembling software applications from prefabricated parts is as old as software engineering itself (McIlroy presented a paper on "Mass-Produced Software Components" at the NATO conference that coined the terms software engineering and software crisis [1]). However, despite significant research effort put into reuse in the 1980s and 1990s, McIlroy's vision has remained stubbornly elusive. Over the years there were many reasons why fine-grained component reuse has failed to take off, but generally speaking there have been three main barriers [2] –

1. there simply were not enough "good" components around to make reuse worthwhile,
2. the recall and precision of the retrieval technologies used to find suitable components was not sufficient,
3. the overall risk and effort involved in finding and evaluating components for reuse was too high compared to the risk and effort involved in building them from scratch.

Over the last few years there have been dramatic improvements with respect to the first two of these. The rapid growth in freely available, open source software repositories such as SourceForge and Google Code as well as the emergence of dedicated search engines that index them (such as Google Code Search, Krugle and merobase) now provide developers with easy access to vast swathes of reusable software. However, these advances have only partially alleviated the third problem. They are necessary but not sufficient. Although the precision of some of the new generation of code search engines is much higher than before [3], the ratio of suitable to non-suitable components in search results is still relatively low and developers still have to evaluate them all by hand. The costs and risks involved in "manual" reuse by directly interacting with code search engines therefore still typically outweigh the benefits.

To fundamentally change the risk/cost/benefit balance and make fine-grained component reuse the rule rather than the exception Garcia et al [4] argue that component search facilities need to be integrated into a fully fledged software reuse environment. Such an environment should (a) allow reuse recommendations to be driven by a background agent that monitors the work of the developer and triggers searches "proactively" (b) provide automatic assistance for query formulation to bridge the gap between the described functionality of a component and the described needs of the developer and (c) make reuse as non-intrusive as possible so that the developer is barely disturbed from his normal work.

We believe that integrating search functionality seamlessly and unobtrusively into standard development environments is only one half of the solution, however. We also believe it is important to substantially raise the quality of research results. Even if component search functionality is offered in a highly unobtrusive way, it will still not be used unless there is a reasonable likelihood that the effort and risk involved in evaluating a component will be worthwhile. We believe the best way to enhance the quality of the results is to exploit the fact that code, unlike most other documents indexed by text-driven search engines, is executable. This means that a component's "fitness for purpose" can be established by testing it. This presupposes the existence of test cases that can be used to test components, but fortunately the trend in modern development approaches such as agile development is to develop test cases before writing code.

In this paper we present a tool, Code Conjurer, (www.code-conjurer.org), that implements these features using the merobase software component search engine (www.merobase.com).

## 2. Proactive Reuse Recommendation

Several prototype tools have been developed to provide assistance to developers based on information garnered from code repositories. The chief examples include Rascal [5], Prospector [6], ParseWeb [7] and Strathcona [8]. These help developers to work out what methods to call in what sequences or provide examples of previous ways in which a component has been reused. However, none is directly focused on finding reusable components and none provides support for proactive recommendations.

The first tool to offer proactive help to users based on information garnered from a code repository was CodeBoker [9]. This tool was focused on reusing good design and coding practices rather than fully blown components per se, but it pioneered the notion of "proactive recommendation". The main weakness of CodeBroker is that it requires components to be "annotated" by developers and is unable to handle normal software modules. Finally, CodeGenie [10] is an Eclipse plug-in that focuses on finding reusable components. However, it is not proactive and requires developers to manually test all reuse candidates locally in their development environments.

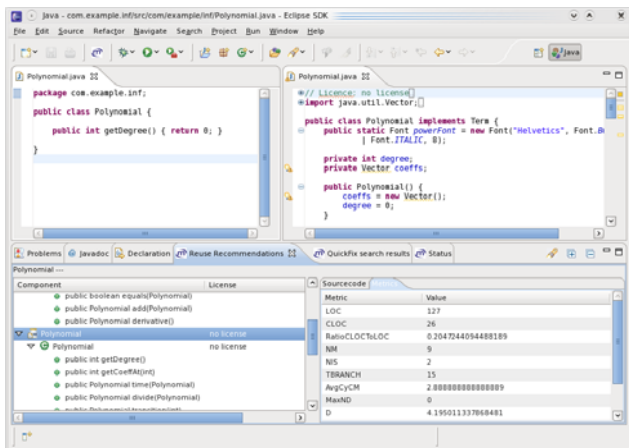

Figure 1: Proactive reuse

At the University of Mannheim we have been working on a plug-in known as Code Conjurer [10] that uses the merobase code search features to realize a software reuse environment of the kind envisaged by Garcia et al. [4] within the Eclipse framework. When plugged into the standard Eclipse Java environment, it allows searches to be initiated from various kinds of Java code fragments at the click of a button. When set into "proactive" mode the plug-in also provides proactive reuse recommendations. It includes an agent that monitors the component under development (CUD) and autonomously recommends potentially interesting candidates for reuse in a non-intrusive way.

When the background agent discovers a significant change in the CUD's interface-defining part (e.g. a method has been added, changed or removed) it triggers a search via the Merobase API. The component is analyzed, its interface is extracted, an MQL (Merobase query language) query is created and user-defined constraints are added (e.g. duplicate filtering or exclusion of interfaces). The resulting list of components is presented to the developer in an Eclipse view, as shown in the bottom left of figure 1. He can then study the components in more detail, review the implemented methods and compare components using different metrics.

If the developer decides that a component is worth reusing, by a simple double-click he can either weave it into the current project, thus overwriting his own code, or can put it into a new project. When the component is inserted into a new project, Code Conjurer automatically detects unresolved dependencies and tries to automatically resolve them (provided that this functionality is activated in the preferences).

During development it may also happen that the component under development needs a new kind of object (e.g. when writing an address book a *Person* object might be necessary). In this case the developer can simply specify the object (e.g. with `Person p = new Person();`) which will lead to an error message displayed by Eclipse indicating that the type cannot be resolved. Using the QuickFix feature of Eclipse, the developer can easily get Code Conjurer to search for a *Person* component and afterwards directly add it to his project thereby avoiding self-development.

Even if the developer does not wish to use one of the recommended components, Code Conjurer provides potentially interesting information about the "typical" or "average" form of the discovered components. Using various clustering techniques, the recommended components are analyzed and a characteristic group picture is created. This information indicates the typical set of methods offered by components matching the developer's partially defined interface. For example, suppose the developer is working on a class *Polynomial*, Code Conjurer can indicate that classes of this name typically offer the following methods :

```
public class Polynomial {
    Polynomial add(Polynomial arg1) {     }
    String toString() {}
    int getDegree() {}
}
```

In contrast to the software reuse environment envisaged by Garcia et al. [4] which only foresees automatic help in query formulation, Code Conjurer extracts all necessary information automatically from the CUD and creates the search queries itself without user involvement. Moreover, the developer does not have to write code according to any particular standard or worry about interacting with the search engine but can fully concentrate on developing his application.

# 3. Test-Driven Software Reuse

Code Conjurer seamlessly and unobtrusively integrates search functionality into the Eclipse development environment, but this still does not ensure that it will be reused in practice. As mentioned above, the value of a software reuse environment is significantly diminished if the quality of the search results is low and the developer has to spend a lot of time "manually" studying and evaluating components. As Mili and Mittermeir pointed out in 1998 [12], software artifacts are not textual documents but are executable modules with observable behavior. Thus, it is possible to test their "fitness for purpose" by checking whether they pass one or more tests. The merobase search engine has therefore been enhanced to support test-driven as well as standard search mechanisms [13]. As soon as an executable test has been defined by the developer and a search is initiated, Code Conjurer sends a request to the merobase server to find matching components that pass the test. Merobase then initiates the test-driven searching process, which in previous papers we have referred to as Extreme harvesting [12] of its synergy with Extreme Programming. This involves the following mains steps -

1. establishing what interface the test is for,
2. performing a normal search on this interface, and
3. testing the results against the provided test-case to filter out those components that match.

The resulting set of component recommendations is of much higher value than one typically generated by regular interface-based matching alone, because the components do what the developer has specified - that is, they pass the provided test-case. Suppose, for example, that the developer needs a *Matrix* component for his application. He starts by writing a test-case for a *Matrix*, specifying all the desired functionality. Code Conjurer then sends this to merobase where the interface defining part of the component is extracted and candidates are identified. In a secured virtual environment, these are then tested against the test-case and only those that match are returned to the developer's IDE as shown in Code Conjurer's recommendations view in the bottom left of figure 2. The components can be directly weaved either into the current developer's project or into a separate one.

It may happen that the initial test is only partially complete so that the components in the recommendations view are a superset of the components that are ultimately of interest to the developer once he has finished writing test cases. Nevertheless, at an early point in the process of writing test cases the developer may already be interested in a group picture of what methods components of the kind he is writing generally implement. Code Conjurer allows him to explore the characteristic group picture and write tests accordingly. This kind of reuse is already mentioned in [8] and known as glass-box reuse.

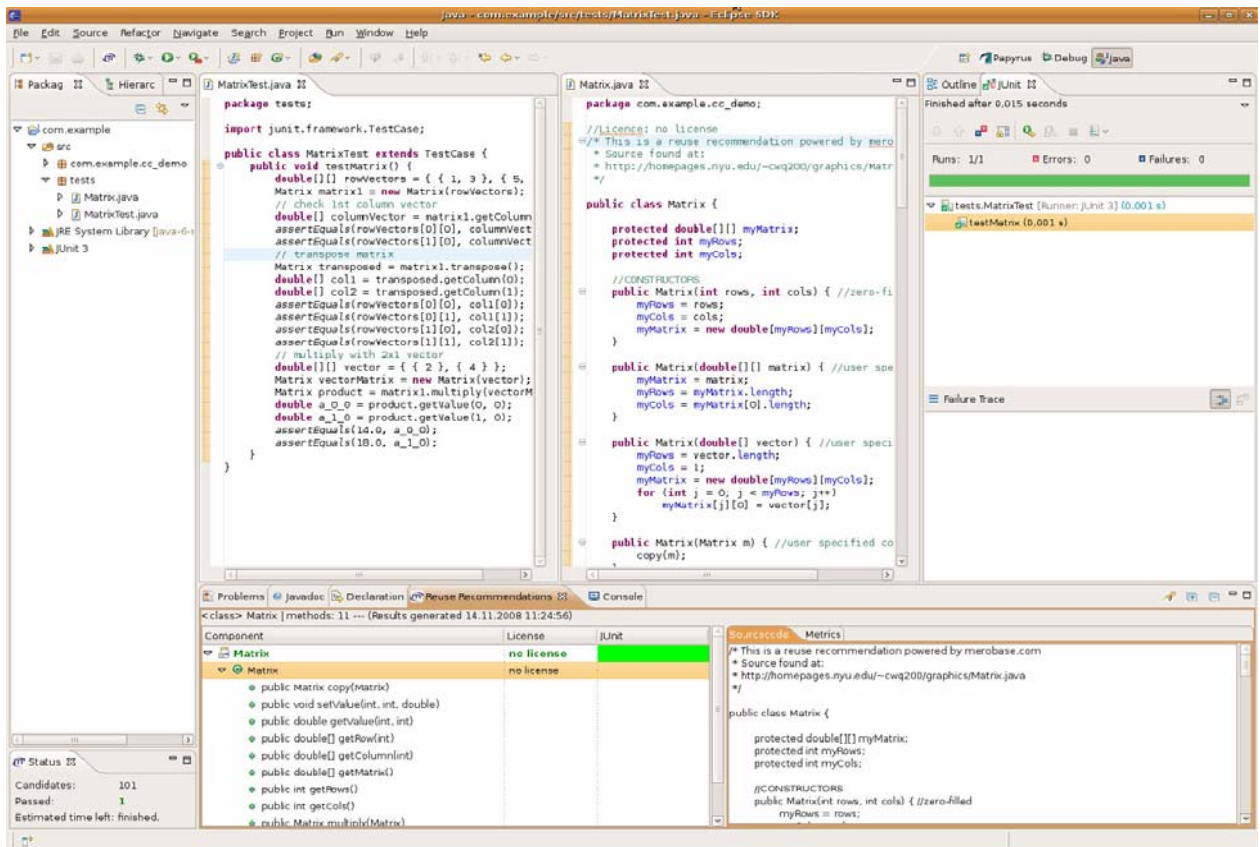

**Figure 2: Test-driven reuse recommendation for a Matrix component**

## 4. Conclusion

We believe that tools like Code Conjurer that combine *proactive reuse recommendation* with *test driven reuse* can for the first time significantly tip the risk/cost/benefit tradeoff between "reuse" versus "build" towards the "reuse" option. The two technologies are also highly synergistic and complement each other's weaknesses. The latter complements the former because it significantly enhances the quality of the search results. In fact the precision of the results from test-driven search is theoretically 1 (the recall is hard to estimate [3]) because all returned components are guaranteed to fulfill the developer's functional requirements as defined by his test case. The former complements the latter because it hides the relatively long search times and low success rates of test driven search. If a developer is not even aware that a test-driven search is being performed on his behalf, the time taken or success ratio is of no concern to him. Any potentially reusable components that the tool is able to "conjurer up" are simply seen as a bonus.

Code Conjurer also provides various other helpful features. For example, component recommendations are generally accompanied with metrics information like the LOC, cyclomatic complexity or Halstead metrics so that non-functional properties of components can be evaluated (bottom right window of figure 1). As well as finding normal functional components Code Conjurer can also find reusable test cases as well. When a developer starts writing a test, Code Conjurer can look for previously indexed tests and offer these for reuse. These can be inspected to give the user an impression of what tests are generally written for a component similar to that he is developing (glass-box reuse), or they can be weaved directly into the developed project and extended or changed as necessary. Our tool also has a dependency resolution feature which analyzes selected components with respect to unresolved dependencies and tries to resolve them using several heuristics – from fast and simple ones to more sophisticated ones. If it finds the needed components, it automatically incorporates them at the necessary places so the error messages of Eclipse disappear.

To conclude, we believe that Code Conjurer, driven by merobase, fulfills the basic vision of a software reuse environment out forward by Garcia et al. [4]. Nevertheless the technology is only just scratching the surface of the development support that can be offered by tools driven by code search engines, and once the remaining possibilities are elaborated we believe the technology will open up a whole new paradigm of search-driven reuse.